\documentstyle{mn}

   \title[Evolutive and composite model for AGNs]
         {Evolutive unification in composite Active Galactic Nuclei
}

\author[L\'{i}pari \& Terlevich]
   {S. L\'{i}pari$^{1}$ and R. Terlevich$^{2,3}$ \\
$^{1}$ C\'ordoba Observatory and CONICET, Laprida 854, 5000 C\'ordoba,
Argentina.\\
$^{2}$ Institute of Astronomy, Madingley Road, Cambridge CB3 OHA.\\
$^{3}$ Instituto Nacional de Astrofisica y Optica (INAOE), Puebla, Mexico.
}

\date{Received     ;
      in original form }

\pagerange{\pageref{firstpage}--\pageref{lastpage}}
\pubyear{2005}

\begin{document}

\maketitle

\label{firstpage}

\begin{abstract}

In this paper we explore an evolutionary Unified scenario involving super
massive black hole (SMBH) and starburst (SB) with outflow, that seems
capable of explaining most of the observational properties
--of at least part-- of AGNs

Our suggestion is explored  inside the expectations of the Starburst model
close associated with the AGN where the narrow  line region (NLR), broad
line region (BLR) and broad absorption line (BAL) region are produced
in part by the outflow process with shells and in compact supernova remnants
(cSNR).

The outflow process in BAL QSOs with extreme IR and  Fe\,{\sc ii} emission
is studied.
In addition, the Fe\,{\sc ii} Problem
regarding the BLR of active galactic nuclei (AGN) is analysed.
Neither the correlations between the BAL, IR emission,
Fe\,{\sc ii} intensity and the intrinsic properties of the AGN are
clearly understood.
We suggest here that the behaviour of the BAL, IR and Fe\,{\sc ii} emission
in AGNs can be understood inside an {\bf evolutionary and composite}
model for  AGNs. 

In our model, strong BAL systems and Fe\,{\sc ii} emission are present
(and intense) in young IR objects. 
Parameters like BALs, IR emission, Fe\,{\sc ii}/H$\beta$ intensity ratio,
Fe\,{\sc ii} equivalent width, broad lines line width,
[O\,{\sc iii}]$\lambda$5007 \AA\ intensity and width, narrow line region
(NLR) size, X-Ray spectral slope in radio quiet AGN plus lobe separation,
and lobe to core intensity ratio in radio loud AGN are proposed to be
fundamentally time dependent variables.
Orientation/obscuration effects take the role of a second parameter
providing the segregation between Sy1/Sy2 and BLRG/NLRG.

\end{abstract}


\section{INTRODUCTION}\label{intro}

Recent work: (i) has {\it provided} strong evidence for SMBH in the
centres of most/all massive bulges; (ii) has {\it proved} the presence
of massive SB in the central regions of a substantial fraction of AGN
[both radio quiet (RQ) and radio loud (RL)].

From 3D and 1D Spectroscopic studies we have found
kinematical and morphological evidence of a close relation
between the AGN and extreme starburst, in nearby ``BAL + IR +
Fe\,{\sc ii} QSOs and mergers" (L\'{i}pari et al. 2005a,b,c,d, 2004a,b,c,d,
2003, 2000, 1994, L\'{i}pari 1994).

In particular, some of the results obtained for \emph{nearby BAL QSOs},
such as strong IR and Fe\,{\sc ii} emission,
strong blue asymmetry/OF in H$\alpha$, radio quietness, and
very weak [O\,{\sc iii}]$\lambda$5007 emission
(Low et al. 1989; Boroson \& Meyers 1992; L\'{\i}pari et al. 1993, 1994,
2003, 2005a; L\'{i}pari 1994; Turnshek et al. 1997),
 can be explained in the framework of the starburst + AGN scenario.
In our study of Mrk\,231 and  IRAS\,0759+6559
(the nearest extreme BAL + IR + GW + Fe\,{\sc ii}  QSOs), we
detected typical characteristics of  young-starburst QSOs.
In our evolutive model for young and composite IR QSOs (see for
references L\'{\i}pari 1994; L\'{i}pari et al. 2005a) it is suggested that
some BAL system plus IR and Fe\,{\sc ii} emission could be linked
to violent super massive--starburst + AGN which can lead to a large-scale
expanding super giant shells, often obscured by dust.
Several articles suggested that this evolutive model
shows a good agreement  with the observations (see Egami et al. 1996;
Canalizo \& Stockton 1997; Lawrence et al. 1997; Canalizo et al. 1998).

In this paper we explore an evolutionary Unified scenario involving both
SMBH and SB with OF, that seems capable of explaining most of the
observational properties --of at least part-- of AGNs

\subsection{Static Unification Model for AGNs}\label{suagn}

The so-called AGN unified models represent an attempt to explain the variety 
of sub-types as due solely to differences
in the orientation of the central object and/or its nearby environment. 
There are two branches of unification, the radio-quiet and the
radio-loud. 

The radio-quiet unification interprets type 2 nuclei, i.e. those dominated by narrow
emission lines, as normal 
nuclei or QSOs whose central regions, containing the ionising source and the
surrounding Broad Line Region (BLR), are obscured by an edge-on opaque dusty
torus. The observed narrow emission lines are produced at large distances
from the nucleus
and result from the ISM being photoinized by the nuclear UV radiation
leaking trough the poles of the dusty torus. 
Reflection of the nuclear light by electrons and/or dust in the extra nuclear
regions provides in some cases a periscopic view into the hidden active
nucleus. In this scenario, normal type 1 nuclei are those seen directly
without obstruction by the dusty torus.

Support for this scenario comes from the discovery of broad permitted lines
in the polarized spectra of nearby type 2's, strengthening the idea that
Seyfert 2's harbour obscured BLR's (e.g., Antonucci \& Miller 1985;
Lawrence 1987, 1991; Antonucci 1993; Miller \& Goodrich  1990).
Moreover `ionization
cones' have been found in several nearby AGNs (e.g., Wilson, Ward \& Haniff
1988; Pogge 1988a,b; Tadhunter \& Tsvetanov 1989; Storchi-Bergmann, Wilson \&
Baldwin 1992). These cones are generally aligned with the radio axis and
perpendicular to the polarization vector (Antonucci 1993).
These findings agree with the unified model, which
predicts that only gas within a biconic region (defined by the opening angle
of the obscuring torus) directly `sees' the nuclear ionizing continuum,
whereas the observer's line of sight to the nucleus is blocked and only a
fraction of the central source's continuum is actually scattered towards
him/her. Elvis \& Lawrence (1988) found that NGC 1068 has a hard X-ray
spectrum typical of a Seyfert 1, with very little X-ray absorption. They
concluded that these findings are consistent with all the observed X-rays
coming from reflection of the hidden Seyfert 1 spectrum by electrons in a
photoionized region, extending the reflection model to the X-ray band (see
however Wilson et al. 1992). 

There are  at least two or three main problems with this simple unification
scenario,

\begin{enumerate}

\item
It does not explain the origin of the strong UV/blue
continuum observed in most type 2 and their low polarization 
(Cid-Fernandes \& Terlevich 1995). Cid-Fernandes \& Terlevich
suggested that active star formation close to the nucleus will explain
this points plus the large strength observed in the near IR stellar lines
of the nuclear spectra (Terlevich, Diaz \& Terlevich 1990). 
The presence of young stars in the nuclear region
of type 2 Seyferts has been confirmed by the detection of absorption lines
from massive star in the UV (Heckman et al. 1995) and IR (Oliva et al. 1995). 

\item
There is a strong possibility that some type 2's are not
obscured type 1's. Several low luminosity AGN are known to have
undergone type transitions from type 1 to type 2 and/or vice-versa (see
Aretxaga \& Terlevich 1994 for a compilation of cases). Such nuclei are
clearly not obscured type 1s. This fact indicates that short term evolution 
is playing an important role at least in some low luminosity AGNs.

\item
Only few high-luminosity type 2 are known with M$_B < -23$  (Osterbrock 1993).

\end{enumerate}

In the radio loud unification core dominated radio sources, i.e. BLRG and
quasars, are systems with pole-on toroids and thus viewed approximately
along the axis of the radio jet with the most extreme cases, i.e. those
where the jet points roughly in the observers direction, showing
superluminal motions in the inner parts of the jet.
Lobe dominated radio galaxies, NLRG, are those systems where
the torus is almost edge-on and the jet is close to the plane of the sky.
There is again convincing evidence for the geometrical unification. 
The R parameter that measures the core to lobe intensity ratio is related
to the FWHM of the Balmer lines.
Lobe dominated radio sources (small R) tend to have broad Balmer lines
while core dominated ones (large R) show only relatively narrow
Balmer lines.

In all, the simple UM
represents a step forward towards the
unification of radio loud AGN. But there are some problems summarized by
Gopal-Krishna, Kulkarni \& Wiita (1996), namely 
the crisis of the relative sizes of radio galaxies and quasars,
the apparent increase of the torus aperture with AGN luminosity.

In addition to the above problems with the simple unified model (UM), 
there are other that affect both the RQ and RL
AGN. These problems are related to the observed behaviour of the permitted
optical Fe\,{\sc ii} emission. 
The Fe\,{\sc ii} emission was first identified in
1966 (Wampler \& Oke 1967) in the spectrum of 3C~273. 
Pioneering work by Steiner (1981) indicated the importance of Fe\,{\sc ii}
emission in AGN and suggested a classification scheme based on the strength
of Fe\,{\sc ii}.
Boroson \& Oke (1984) and Boroson, Persson \& Oke (1985) studied the
nebulosity around high luminosity quasars and found a clear segregation two
groups according to the intensity of the NLR, the strength of Fe\,{\sc ii},
radio morphology and spectral index and Balmer line width.
During the last decade considerable effort has been devoted to understand
the origin of Fe\,{\sc ii} optical emission observed in many Seyfert 1 and
QSOs.
Extreme Fe\,{\sc ii} emission is not reproduced by standard photoionization
models which cannot even account for the observed Fe\,{\sc ii} 4570 \AA
/H$\beta$ line intensity ratios that is typically $>$6 but as large as 30
in some extreme emitters.
This apparent failure of the standard photoionization model has lead
to the search of correlations between Fe\,{\sc ii} intensity and other AGN
properties.

There is further information apart from the optical and radio properties
that show tantalizing and unexplained trends between observables.
A correlation between the slope of the X-ray spectrum and the equivalent
width of the Fe\,{\sc ii} $\lambda$ 4570 \AA\ emission was found for
samples of QSOs (Wilkes \& Elvis 1987) indicating that strong Fe\,{\sc ii}
emitters are the AGNs with the softest X-ray spectrum.
This is very puzzling  because if, as suggested by some authors,
the strong  Fe\,{\sc ii} emission is produced by the presence of zones
of the BLR of very high opacity to ionizing radiation, thus penetrated
only by very hard X-rays, the opposite correlation should be found.
Also puzzling, is that the most extreme Fe\,{\sc ii} emitters are
radio-quiet quasars; contrary to what would be expected if the jets
responsible for producing compact radio sources,
are also to produce the strong Fe\,{\sc ii} emission. 

But perhaps the most intriguing discovery is the strong anti-correlation 
found by Boroson \& Green (1992)
between the Fe\,{\sc ii} 4570 \AA /H$\beta$ line ratio and the peak intensity of
[O\,{\sc iii}] 5007 \AA.
Strong Fe\,{\sc ii} emitters have weak or absent [O\,{\sc iii}] while weak
Fe\,{\sc ii} emitters have the strongest [O\,{\sc iii}] emission. 
Because the [O\,{\sc iii}] should be practically
orientation-independent this anti-correlation is an indication 
that the nucleus of these galaxies are not seen along a preferred line of 
sight (i.e., this relation can not be explained by orientation effects).
Furthermore, using only geometrical/orientation based arguments 
is very difficult to explain the large dynamical range observed in
the [O\,{\sc iii}]$\lambda$5007/H$\beta$ ratio. This ratio ranges from values larger 
than 20 in type 1 objects with weak or absent Fe\,{\sc ii} emission to
essentially no detection of [O\,{\sc iii}]$\lambda$5007
in nuclei with strong Fe\,{\sc ii} emission.

Boroson (2002) extends the sample to include radio-loud AGN. His analysis
shows a clear separation in the plane PC1-PC2 between radio loud and
radio quit.
His proposal is that BH mass and ANG luminosity are the key distinctions
between the different class of objects.

\subsection{Evolutive Unification Model for AGNs}\label{euagn}

Some connection between galaxy/star formation and nuclear activity has been
suggested by several researchers (among others: Terlevich \& Melnik
1985, Perry \& Dyson 1985, 1992; Norman \& Scoville 1988; Terlevich
\& Boyle 1993). 
Several works in the QSO luminosity function tend to underline the possible
close relation between nuclear starbursts and AGN (e.g., Terlevich \& Boyle
1993; Hoehnelt \& Rees 1993).
In luminous IR mergers and IR QSOs we proposed a composite scenario
for the origin of the BAL and IR + Fe II emission (see for details
Lipari et al. 2005a, 2003, 1994, 1993; Lipari 1994; Lipari \& Macchetto 1992;
Sanders et al. 1995, 1988a,b; Joseph \& Wright 1985).

We propose a scenario where a SMBH is formed during the formation 
of the galactic core and an AGN powered by accretion. Which is formed
by the interaction between a nuclear SB and the SMBH.
The evolution of the SB generates a diversity of intrinsic properties
that when combined extrinsic ones such a the orientation,
are able to explain most of the observed differences between
the different types of AGN

It seems that the problems of the static UM derive from
the fact that it is static i.e. does not consider the evolution of the
AGN itself on time scales less than $10^8$yr.
We propose that intrinsic parameters like the angular size of radio sources
and their radio luminosity, the  BLR spectrum (emission line width,  
Fe\,{\sc ii} intensity, etc), the NLR spectrum, NLR size and NLR luminosity, 
all depend on the evolutionary age of the AGN, while the observed
parameters depend on age and also on the orientation of the AGN and 
its environment with respect to the observer.

The timescale for the development of megapersec size radio sources 
is similar to the lifetime of the AGN (Fanti et al. 1995). 
The age or lifetime of the AGN should also play an important role
(Gopal-Krishna et al. 1996).

We propose a bi-parametric   {\bf evolutionary} model for AGNs.
Intrinsic parameters like the BAL,  Fe\,{\sc ii} intensity and BLR
spectrum, NLR size and luminosity and radio luminosity, size and
morphology, all evolve with a time scale of less than $10^8$ yr.
Young AGN are obscured BAL and strong Fe\,{\sc ii} emitters
with relatively narrow line BLR and a compact and faint NLR; their radio
emission is also compact.
Old AGN are weak Fe\,{\sc ii} emitters with broad line BLR and extended
and bright NLR and fully developed radio lobes.
The orientation of the AGN/toroid takes the role
of a second parameter providing the segregation between type 1/type 2,
and BLRG/NLRG.

Radio emission comes later in the evolution, after the action of the 
winds from the SB have removed the ISM towards the regions of
larger density gradient in the circumnuclear ISM distribution.
Also the SB ejecta action may help in regularize and order
the central magnetic field. Those systems were the spin of the
SMBH points in the direction of the SB ejecta are the ones with higher
probability of becoming powerful radio sources.

\section{The Evolution from mergers to QSOs}
\label{evoleqso}

\subsection{Evolutive links between IR Mergers and IR QSOs, with outflow}
\label{linkmergerqso}

The luminosities and space densities of luminous IR galaxies (LIRGs;
L$_{IR} > 10^{11} L_{\odot}$)
in the local Universe are similar to those of
quasi-stellar objects (QSOs; Soifer, Houck \& Neugebauer 1987).
In addition, at the highest IR luminosities, the presence of AGNs (and
mergers) in LIRGs becomes important.
Thus LIRGs probably represent an important stage
in the formation of QSOs and elliptical galaxies.
These results strongly suggest that it is important to
perform detailed studies of possible links among mergers, LIRGs, QSOs
and elliptical galaxies (see for references L\'{i}pari et al. 2005a)

In addition, the  detection of a  correlation  between the mass of
galactic bulges and the mass of supermasive black hole is a confirmation
that the formation and evolution of galaxies (bulges/ellipticals, mergers)
and super massive black hole (AGNs and QSOs) are physically related to
one another (Magorrian et al. 1998; Ferrarese \& Merrit 2000; Gebhardt et al.
2000; Kormendy 2000; Merrit \& Ferrarese 2001; Kormendy \& Richstone 1995).

In the last years, several possible \emph{links} between \emph{mergers,
starbursts, IR~QSOs and ellipticals} have been proposed. Specifically,
Joseph et al., Sanders et al. and L\'{\i}pari et al. suggested three
complementary sequences and evolutive--links:

(i) merger $\to$ giant shocks $\to$ super-starbursts + galactic
winds $\to$ elliptical galaxies;

(ii) merger $\to$ H$_2$-inflow (starbursts) $\to$ cold ULIRGs
$\to$  warm ULIRGs + QSOs;

(iii) merger{\bf /s} $\to$  extreme starburst + galactic-wind
(inflow + outflow) $\to$  IR + Fe\,{\sc ii} + BAL composite/transition QSOs
$\to$ standard QSOs and ellipticals $\to$ ? (galaxy remnants).

In this sequence a main step is not -yet- explored/studied. Which is
the end phases of the evolution of the host galaxies + QSOs. Recently,
we started theoretical studies (L\'{i}pari \& Terlevich 2006, in
preparation) for one of the more simple case: the end phase of 
elliptical galaxies. In particular, for this type of galaxy we are studying
the processes when: the amount of gas is very low, and the star
formation and SN events are almost finished.
At this stage we can observe mainly stars+SN remnants and SMBH, which are
very compact and dark objects.

\subsection{The Evolution in the IR colour--colour diagram of
mergers and QSOs (with outflow/BALs)}
\label{discussion-trannsi2}

The IRAS colour-colour diagrams have been used as an important 
tool to detect and discriminate different types of activity in the 
nuclear/circumnuclear regions of galaxies.
L\'{i}pari (1994) already found that the IR colours (i.e., IR energy
distribution) of  $\sim$10 extreme  IR +
Fe {\sc ii} QSOs are distributed between the power law (PL) and the
black-body (BB) regions: i.e., the {\it transition area}.
On the other hand, the  low and moderate IR--Fe\,{\sc ii} emitters
are  located mainly in the PL region. 
In particular, we detected that  Mrk 231 and IRAS~07598+6508 
(the nearest IR + GW/OF + Fe {\sc ii} + BAL QSOs) have a
close position in this diagram: near
to the BB area; thus showing both systems strong starburst components.
Recently, Canalizo \& Stockton (2001) confirmed that the host galaxies of
both QSOs have strong starburst populations (using Keck spectroscopy).

It is important to remark
that of a total of $\sim$10 IR transition objects of this original
sample, the first 4 systems are BAL IR QSOs. Therefore, we
already suggested that BALs IR QSOs (like Mrk 231, IRAS~07598+6508,
IRAS 17002+5153 and IRAS 14026+4341) could be associated with the {\it
young phase of the QSO activity}.

Very recently, using our data base of more than 50 IR Mergers and QSOs
with galactic winds and using for comparison the large sample of standard
PG QSO (from Boroson \& Green 1992) we have expanded our previous study.
Figure 1 (adapted from L\'{i}pari et al. 2005a: their Fig. 15) show the IR
energy distribution [spectral indexes
$\alpha$${(60,25)}$ vs. $\alpha$${(100,60)}$; where $\alpha$${(\lambda 2,
\lambda 1)}$ = --log[F($\lambda 2$)/F($\lambda 1$)]/log[$\lambda 2$/$\lambda 
1$]] for:
(i)  IR mergers and IR QSO with GW (originally 51 IR systems);
(ii) standard QSOs from the PG QSOs sample of Boroson \& Green (1992;
originally 87 PG QSOs that have $z\leq$0.5).

IR fluxes--densities, in the bands of 12, 25, 60 and 100 $\mu$m, were
obtained from the IRAS and ISO Archival Catalogue (using NED).
Only objects with a good detection in the three required bands have been
included. Also, the localisation of the three main regions in 
this colour-colour diagram (i.e., the QSOs/Seyferts, Starbursts, and
powerful IR galaxies areas) have been plotted.
An inspection of this diagram clearly shows the following:

\begin{enumerate}

\item
All the IR mergers with low velocity OF (LVOF) are located very close to
the BB and starburst area.

\item
Almost all the IR QSOs with extreme velocity OF (EVOF) are located in the
transition region.

\item
The standard QSOs and radio QSOs are located around the PL region.

\item
All the BAL IR QSOs are located in the transition region, in almost a clear
sequence: from
Mrk 231 (close to the BB area) $\to$  IRAS 07598+6508 $\to$
IRAS04505--2958 $\to$ IRAS 21219-1757
$\to$ IRAS/PG 17072+5153 and IRAS 14026+4341 (close to the PL area) $\to$
standard QSOs.

\end{enumerate}

These results first confirm our previous finding (obtained from
a small sample of IR galaxies): in the sense that
{\it IR QSOs are probably ``young, composite and transition" objects}
(between IR mergers and standard QSOs).
Furthermore, in this IR colours diagram a main evolutive parameter
is the values of the OF: from mergers with LVOFs to OSOs with EVOFs.

\section{The Young and Composite Nature of BALs + IR +
Fe\,{\sc ii} mergers/QSOs}
\label{bal}

\subsection{The BAL Phenomenon}\label{euagn}

In the last decades, the nature of the BAL phenomenon and its relation to
the overall quasar population has been the subject of debate
(see for references Lewis, Chapman \& Zuncic 2003).
Two main interpretation for the occurrence of BALs are proposed:
 the orientation and evolution hypothesis.
According to the orientation interpretation  "all QSOs" possess
BAL OF, so that the frequency of detection only translate to the rate at
which our line of sight intercept the OF (Weyman et al. 1991).
According to the evolution hypothesis the rate of the BAL phenomenon is
interpreted as a particular phase of the QSO's life (L\'{i}pari 1994;
Voit, Weymann \& Korista 1993).

Observational evidence supporting the orientation hypothesis come from
spectral comparison of BAL and non-BAL QSOs (Weyman et al. 1991) and
polarization studies (Hines \& Will 1995; Goodrich \& Miller 1995;
Schmidt \& Hines 1999).

Evidence in favour of the evolution hypothesis comes largely
from the high number of BALs detection in IR + Fe II QSOs/mergers
(see for references L\'{i}pari et al. 2005a, 2003, 1994; L\'{i}pari 1994;
Boroson \& Meyer 1992; Low et al. 1988, 1989).
Further support for the evolution hypothesis has been provided for
radio observations of BAL QSOs, which are inconsistent with
{\it only orientation schemes} (Becker et al. 2000, 1997).

In the last decade, we proposed and evolutive scenario for
BAL + IR + Fe\,{\sc ii} QSOs, where mergers fuel extreme star formation
processes and  AGNs, resulting in strong dust and IR emission, large
number of SN and Hyper Nova events with expanding super giant
bubbles and shell. The BALs in IR + Fe\,{\sc ii} QSOs were associated with
this composite nature of these systems (L\'{i}pari et al. 2005a; L\'{i}pari
1994).

\subsection{BALs in IR + GW + Fe\,{\sc ii} mergers/QSOs} \label{programme2}

In general,
the role of BALs in IR+GW/OF+Fe {\sc ii} QSOs/mergers must be
carefully considered, since: (i) Low et al.
(1989) and Boroson \& Meyers (1992) found that IR selected QSOs
show a 27\% low-ionization BAL QSO fraction compared with 1.4\%
for the optically selected high-redshift QSOs sample (Weymann et
al. 1991); (ii) extreme IR galaxies (ULIRGs) are mainly mergers
(see Section 1); (iii) recently Maiolino et al. (2003) reported
also a high fraction of BAL QSOs at very high redshift (z $\sim$6).
The high percent of occurrence of broad absorption in extreme
IR + GW/OF + Fe {\sc ii} QSOs/mergers may be signals a fundamental relation
(rather than merely a coincidence).

L\'{\i}pari et al. (1993, 1994, 2003, 2004a,d, 2005a); Scoville \&
Norman (1995);
Perry 1992; Perry \& Dyson 1992; Dyson et al. 1992;
Egami et al. (1996); Lawrence et al. (1997) and others proposed that
the extreme BAL + IR + GW/OF + Fe\,{\sc ii} phenomena are related --at least
in part-- to the {\it end phase of an ``extreme--starburst + AGN" and the
associated powerful bubble/galactic--wind}. At the final stage of a
strong starburst, i.e., type II SN phase
([8-60]\,$\times10^{6}$\,yr from the initial burst; Terlevich et
al. 1992; Norman \& Ikeuchi 1989; Suchkov et al. 1994) giant
galactic arcs and extreme Fe\,{\sc ii}+BAL systems can appear.

Recently, a new method for decoupling the spectra of the QSO/AGN from the
host galaxy --using 3D spectroscopy-- was developed (Sanchez et al. 2004).
Using this new method for the very deep Gemini GMOS 3D spectroscopic data
 of the nearest BAL + IR + Fe II QSO Mrk 231, we have obtained
the clean spectra of the QSO and the host galaxy for the nuclear region
(r $\sim$ 2$''$).
From this study, the following main result was found:
in the pure host galaxy spectrum a strong nuclear starburst component
was clearly observed (for the first time, at optical wavelength), mainly as
a very strong increase in the flux, at the blue region (Lipari et al. 2006).
This result confirm the composite nature of the very nucleus of Mrk 231.

\subsection{The new BAL + IR  + Fe {\sc ii} QSO  IRAS04505--2958
and the Composite hyper--wind model for the origin of BALs and
Ly$\alpha$ blobs}
\label{I04}

It is important to remark that in the Fig. 1 IRAS 04505--2958 is located
exactly in the sequence of BAL QSOs,
between the positions of IRAS 07598+6508 and IRAS 21219-1757.
This IR + GW + Fe {\sc ii} QSO shows probably the more interesting OF
supershell/arc detected to date.
Which is  very extended (of $\sim$20--25 kpc) and it is located very
far from the nucleus (at r $\sim$ 15 kpc; see for details their Fig. 16a).
The UV HST FOS spectra --of IRAS 04505--2958-- clearly show a BAL system  at
C {\sc iv}$\lambda$1549 emission line (see L\'{i}pari et al. 2005a).

On the other hand, the ionizing radiation from the newly formed young
stars should lead to
prominent Ly$\alpha$ emission due to recombination of the hydrogen in
the ISM. Thus, extended Ly$\alpha$ emission could be an important
spectral signature of young systems, specially at  high z.

In the last years very extended blobs -specially in Ly$\alpha$-  has been
detected in a variety of high (and low) redshift objects:
(i) Several recent surveys of Ly$\alpha$ emitters at high z (Steidel et al.
2000; Keel et al. 1999; Francis et al. 2001;  Matsuda
et al. 2004) have established the existence of extended, highly luminous
Ly$\alpha$ halos (of 50-100 kpc and 1.4 $\times$ 10$^{44}$ erg s$^{-1}$).
Taniguchi \& Shioya (2000) suggested a starburst hyper--wind scenario for
the origin of the Ly$\alpha$ blobs.
(ii) The results of the surveys --at high z-- of bright Sub-mm source
(Chapman et al. 2004a; Bower et al. 2004, Swinbank et al. 2005) suggest that
a very high fraction (3/4) of these sources are extended and complex (i.e.,
showing extended and highly luminous Ly$\alpha$ halos; Chapman et al. 2004a).
(iii) In the last decade, several  extended Ly$\alpha$ halos/blobs were
detected in high redshift radio sources (see for references Reuland et al.
2003). In particular, Reuland et al. (2003) and Dey et al. (1997) proposed
that in high z radio source: starburst-super winds and radiation pressure
from AGN can disrupt and stop the accretion process and to generate the
extended Ly$\alpha$ nebulae/halos.

It is important to remark, that even in the more extended and studied blob:
i.e. LAB1 in SSA 22 (Steidel et al. 2000) there are different
interpretations, about  their origin.
In particular, Matsuda  et al. (2004) proposed the
presence of extended shells and arcs (in this blob/LAB1).
They  associated these shells/blob with a strong star formation process,
which is in agreement with the starburst scenario for the hyper--wind model
(proposed by Taniguchi \& Shioya 2000). However, Chapman
et al. (2004b) suggested that the multiwavelength data of LAB1 are consistent
with an obscured AGN (as the source of the blob).

Very recently, L\'{i}pari et al. (2005a) proposed a {\it composite 
hyper--wind scenario} in order to explain the very extended blob/shell
(of 30 kpc) found in the new BAL QSO IRAS04505-2958 (this BAL IR-QSO was
discovered using the IR colour-colour diagram: Fig. 15 in L\'{i}pari et al.
2005a).
In addition, they suggested  that extreme explosions and extreme starbursts
are associated mainly with the interaction between: the QSO and the nuclear
star formation process.

At high redshift (z $>$ 2.0), we are studying deep 3D spectroscopic data of 
Sub-mm and Radio BAL-QSOs, using Gemini+GMOS and ESO VLT+VIMOS.
It is important to remark, that the link between our two programs
--at high and low z-- is not only about some individual IR BAL-QSOs, since
luminous Sub-mm source at high z imply --in the rest-frame--
luminous IR sources. Thus, probably we are studying the same type of objects
and similar physical processes (in both programs).

In addition,  L\'{i}pari et al. (2004a, 2003) found that 75$\%$ of
IR QSOs/mergers (including BAL QSOs) show clear evidence of OF. Some of these
OFs generate super giant galactic bubbles, and for the case of NGC5514 the
3D spectroscopy maps show that the bubble was detected just in the rupture
and preblowout phase. It is interesting to note that the same percent: of
75$\%$ was found by Chapman et al. (2004b) in their study of Sub-mm sources
showing extended and highly luminous Ly$\alpha$ halos (the Sub-mm sources
are the high-z --or redshifted-- version of luminous IR galaxies).

Therefore, extreme OF associated with jets and giant explosions/ hypernovae
could generate extreme galactic-winds that produce expanding shells,  
which could generate BAL systems and extended blobs or halos (L\'{i}pari et al.
2005a,b,c,d, 2003; Punsly \& L\'{i}pari 2005; Reuland et al. 2003; L\'{i}pari 1994).

\subsection{3D spectroscopy of Mrk 231: the nearest BAL + IR +
Fe\,{\sc ii} QSO}

Very recently, we found for Mark 231 that the BAL I system could be
associated with  bipolar outflow generated by the weak/sub-relativistic
jet; and the BAL III system with  a supergiant explosive events (L\'{i}pari
et al. 2005a; Punsly \& L\'{i}pari 2005).

The variability of the short lived BAL--III Na ID system was studied,
covering almost all the period in which this system appeared (between
$\sim$1984--2004).
We found that the BAL-III light curve (LC) is very similar to the shape of
a SN LC. The origin of this BAL-III system was discussed, mainly in
the frame work of an  extreme explosive event.

The HST images of Mrk 231 show 4 (or possibly 5) nuclear superbubbles or
shells with radius r $\sim$ 2.9, 1.5, 1.0, 0.6 and 0.2 kpc.
For these bubbles, the 3D H$\alpha$ velocity field (VF) map
and 3D spectra show in the 3 more external bubbles (S1, S2, S3), 
multiple emission line components with OF velocities, of 
$\langle V_{\rm OF Bubble}\rangle$ S1, S2 and S3 $= [-(650-420) \pm 30],
[-500 \pm 30]$, and $[-230 \pm 30]$\,km\,s$^{-1}$.
We suggest that these giant bubbles are associated with the large scale
nuclear OF component, which is generated --at least in part-- by the extreme
nuclear starburst: giant-SN/hypernova explosions.

\subsection{The relation between  BAL + IR + Fe {\sc ii} 
young QSOs and very high redshift BAL QSOs}
\label{discussion-transi4}

It has been proposed that  extreme starburst + galactic wind processes
associated with IR mergers could play a relevant role in the formation and
evolution of galaxies and QSOs/AGNs, i.e. in their structure, kinematics,
metallicity, etc. Furthermore,
recent detailed observations and theoretical studies have confirmed that 
OF, galactic winds, BAL, large amount of gas+dust and strong Fe {\sc ii}
emission are important components and processes at high redshift
($z$ $\sim$ 4--6), when the galaxies and QSOs formed
(Frye, Broadhurst, \& Benitez 2002; Ajiki et al. 2002;
Pettini et al. 2001; Dawson et al. 2002; Carilli et al. 2004a,b;
Solomon et al. 2003; Taniguchi \& Sioya 2000; Barth et al. 2003;
Iwamuro et al. 2002; Freudling, Corbin, \& Korista 2003; Maiolino
et al. 2004a,b).

On the other hand,
Maiolino et al. (2004a,b, 2003)  presented near-IR spectra
of eight of the more distant QSO (at 4.9 $<$ z $<$ 6.4). Half of these
QSOs are characterised by strong UV BAL systems (at C IV, Mg II, Si IV, Al
III lines).
Although the sample is small, the large fraction of BAL QSOs suggest that
the accretion of gas, the amount of dust and the presence of OF process
are larger (in these objects) than in standard QSO at z $<$ 4.0.
They also suggested that the very high amount of dust was generated by
early explosions of SNe (Maiolino et al. 2004b).

Finally, it is important to remark the similar properties  found in
IR + GW/OF + Fe {\sc ii} + BAL QSOs at low redshift
and  very high redshift BAL QSOs (at z $\sim$ 6.0; Maiolino et al. 2003,
2004a,b; Carilli et al. 2004a,b). According to these similarities,
 we proposed that the {\it phase of young QSO} could be associated
 with the following main processes:
(i) In young QSOs with extremely large amount of gas (concentrate in their
nuclear region). The accretion rate of gas --by the SMBH-- could be
extremely high (see Maiolino et al. 2004a). 
(ii) In addition this extremely large amount of molecular gas
 could generate extreme starbursts; and the presence of AGNs
 could increase the SF close to the nucleus. Specially in the
 accretion disks, with properties --of the SF-- similar to the
 population III of stars.
In these extreme starbursts --close associated with QSOs/AGNs-- it
is expected  giant SN or HyN explosions.
(iii) In young and distant QSOs the very high number of BAL detections
suggest that composite OFs (or EVOFs) play a main role in
their evolution.

\section{The Fe\,{\sc ii} Problem and the Fe\,{\sc ii}
Correlations}\label{fe2}

The optical spectrum of many AGN is dominated by two broad permitted
Fe\,{\sc ii} emission line blends, one centred at about 4570\AA
(Fe\,{\sc ii}~4570) the other centred at 5350 \AA.
Optical Fe\,{\sc ii} has now been measured in the spectrum of several hundred broad
line AGN, showing a large range of intensity relative to Balmer recombination
lines. 

From the theoretical point of view, considerable efforts have been devoted
to understand the origin of Fe\,{\sc ii} optical emission in broad line
AGN during the last decades.
However, extreme Fe\,{\sc ii} emission is not well explained by standard
photoionization models which cannot account for Fe\,{\sc ii} 4570
A/H$\beta$ line intensity ratios larger than 6 (but it goes from 0 to 30).
The origin of this range of values remains unexplained.

The source of heating for the Fe\,{\sc ii} emitting gas could be associated with:

\begin{itemize}

\item
Zones of the Broad Line Region of very high opacity to ionizing radiation,
thus penetrated only by very hard X--rays (Kwan \& Krolik 1981) 

\item
Jets responsible for the compact radio source (Norman \& Miley 1984). 
However, it should be noted that the most extreme Fe\,{\sc ii}
emitters are radio--quiet quasars while Fe\,{\sc ii} is not detected in many
radio loud AGN.

\item
Nuclear starburst where violent
star formation processes occur in high metallicity/pressure environments 
(Terlevich \& Melnick 1988).

\item
Collisional ionization process (Joly 1987, 1993; Veron et al. 2006), which could
be related mainly with shocks, in out flow events.

\end{itemize}

In order to learn about the main factors controlling the optical Fe\,{\sc ii}
emission, relations have been searched between the optical Fe\,{\sc ii} 
equivalent width --or Fe\,{\sc ii} relative intensity-- and AGN  parameters.
In recent years, there were many reported correlations involving the Fe\,{\sc ii}
emission. Some of them seem to be fundamental to AGNs (Steiner 1981, Boroson
\& Green 1992).
The following are some of the most important relations and the
associated interpretations reported for Fe\,{\sc ii} emitters,

\begin{enumerate}

\item
Among the Fe\,{\sc ii} relations the most important one is probably 
the anti-correlation between   
(Fe\,{\sc ii}$\lambda$4570 / H$\beta$) and
I([O\,{\sc iii}]$\lambda$5007 / H$\beta$) emission line ratios found
by Boroson \& Green (1992)
This anti-correlation represents the first eigenvector in the 
principal component analysis of a sample of more than 100 QSOs with high S/N
spectroscopy,  indicating that represents a fundamental relation.
This unexpected result links the properties of the BLR with those of the NLR.
Why [O\,{\sc iii}]$\lambda$5007 / H$\beta$ (and in general all forbidden
lines) should be weak or even absent in strong Fe\,{\sc ii} emitters or
why AGN with strong forbidden lines have no detectable Fe\,{\sc ii} emission?
Because the [O\,{\sc iii}] should be practically
orientation-independent this anti-correlation is an indication 
that the nucleus of these galaxies are not seen along a preferred line of 
sight (i.e., this relation can not be explained by orientation effects).
Furthermore, using only geometrical/orientation based arguments 
is very difficult to explain the large dynamical range observed in the
[O\,{\sc iii}]$\lambda$5007/H$\beta$ ratio.
This ratio ranges from values larger than 20 in type 1 objects with weak
or absent Fe\,{\sc ii} emission to essentially no detection of
[O\,{\sc iii}]$\lambda$5007 in nuclei with strong Fe\,{\sc ii} emission.

\item
An anti-correlation between I(Fe\,{\sc ii} $\lambda$4570) or 
I(Fe\,{\sc ii} $\lambda$4570/H$\beta$) and FHWM(H$\beta$) has been observed
(Boroson et al. 1985). Zheng \& O'Brien (1990) claim that this
anti-correlation prove that the Fe\,{\sc ii} emission
is more aspect dependent than H$\beta$ or the underlying continuum. However, 
Boroson \& Green (1992) find similar line widths
for Fe\,{\sc ii} and Balmer emission in the Palomar-Green QSO sample making
very difficult to support a geometry in which they do not originate 
in the same region. 

\item
An anti-correlation between EW(Fe\,{\sc ii}$\lambda$4570 ) and {\it soft}
(0.2-3.5 keV) X-ray index has been observed (Wilkes \& Elvis 1987).
Boroson (1989) and Zheng \& O'Brien (1990) refuted this relation,
but Shastri et al. (1993) confirmed Wilkes \& Elvis findings and explained 
the differences with Zheng \& O'Brien (1990)
mainly as a consequence of the relation being significant in radio quiet
AGN only. A related correlation 
between (Fe\,{\sc ii}$\lambda$4570 /H$\beta$ ) and {\it soft}
(0.2-3.5 keV) X-ray index has been confirmed by Laor et al. (1994) recent
study of the X~ray properties of a subset of bright Palomar-Green QSO.

\item
There is trend between IR luminosity and Fe\,{\sc ii} 
in the sense that IR luminous AGN with broad
line emission have strong Fe\,{\sc ii} emission (see Low et al. 1988, 1989;
Boroson \& Mayer 1992).
L\'{i}pari, Terlevich, \& Macchetto (1993) found that almost 
100 \% of the AGN with extremely strong
optical Fe\,{\sc ii} blends are ultra-luminous IR AGN.

A correlation
between Fe\,{\sc ii} emission and the far-IR index $\alpha$(25,60) has been
reported by Keel et al. (1994); they noted that the correlations 
between Fe\,{\sc ii} and far-IR emission present a clear problem to 
unification models based in orientation effects 
(the far-IR luminosity should be almost orientation-independent).

\item
A trend has been suggested between (Fe\,{\sc ii}$\lambda$4570 /H$\beta$)
and R the ratio of the core to extended lobe radio fluxes. R is large
in core dominated radio sources.

\item
Boroson \& Green (1992), found a marked dependence of the broad H$\beta$
asymmetry with Fe\,{\sc ii} strength. Strong Fe\,{\sc ii} emitters tend
to have blue asymmetric lines while weak Fe\,{\sc ii} is associated with
red asymmetry in the form of a red wing extension.

\end{enumerate}

Is not possible to interpret all of these correlations 
inside current unified models (Boroson 2002; Boroson \& Green 1992;
Laor et al. 1994; Joly 1993).
Furthermore, since the discovery of AGN several models have been proposed
to explain their observed line ratios (Shields et al. 1972; Netzer 1976;
Krolik \& McKee 1978; Kallman \& Krolik 1986; Collin-Souffrin 1986;
Collin-Souffrin \& Dumont 1986; Ferland \& Rees 1988). These models
are quite successful in accounting for the narrow lines and many of the
strongest broad lines but failed with the Fe\,{\sc ii} lines that in many
AGN emit a substantial fraction of all the BLR energy.
Current models predict a maximum Fe\,{\sc ii} intensity far below the
observed average value (for a review see Joly 1993).
Also the observed range in Fe\,{\sc ii}$_{TOT}$/H$\beta$ is huge, it
goes from essentially 0 up  to about 30. The origin of this range in
the ratio of two broad permitted lines also remains largely unexplained.

On the other hand, recent work has pointed towards a possible connection
of AGN with strong Fe\,{\sc ii} emission with Starburst activity to
explain at least part of the observed activity. The strong
Fe\,{\sc ii} emitters representing the link between starbursts and
classical or normal Fe\,{\sc ii} AGN (L\'{i}pari et al. 2005a, 2004d,
2003, 1994, 1993).
The prominence of Fe\,{\sc ii} emission in quasars with strong infrared to
optical luminosity ratio (Low et al. 1989) is consistent with the assumptions
of a heating source linked to violent star formation processes (Terlevich \&
Melnick 1985) and of Fe\,{\sc ii} emission by rapidly cooling supernovae
remnants occurring in the central regions of massive galaxies undergoing
intense starburst activity (Terlevich et al. 1992).
In the Starburst model for AGN the BLR is produced in compact SNR (cSNR)
and the observed emission lines are the product of reprocessing of shock
radiation by two high density thin shells and the ejecta (Terlevich et al.
1992). One important aspect of this model is the abundances of the ionized
gas emitting the BLR lines are the abundances of the envelope of the star
and the SN ejecta and not the abundances of the ISM.
Given that the expectation is that the ejecta of massive SN type II is
extremely metal rich with respect to the galaxy ISM, there is a good scope to
produce strong metal emission lines particularly of Fe peak elements in the
associated cSNR.




Boroson (2002) suggests that the physical parameter behind the RL/RQ
segregation is the mass of the BH as evidenced by the H$\beta$ FWHM
difference between the groups. 
The fact that there is an anti correlation between [O\,{\sc iii}]FWHM
and H$\beta$FWHM tend to suggest exactly the opposite.
Whittle et al. have found that FWHM [O\,{\sc iii}] is a good indicator
of the bulge velocity dispersion and therefore its mass.
If the correlation between bulge mass and BH mass holds for this sample
the fact that the mean FWHM [O\,{\sc iii}] in RQ is larger than in RL
systems suggests that they are located in less massive bulges and
therefore are powered by less massive BH.

It is important to remark that L\'{i}pari (1994) and
Lawrence et al. (1997) already proposed that the nuclear OF is a main
process/parameter, that
could explain some of these correlations, observed in AGNs and QSOs.
Furthermore, using our database of IR QSOs with gallactic winds and OF we
found a correlation between the Fe\,{\sc ii}$\lambda4570$/H$\beta$ vs.
velocity of OF (see Figure 2; which is adapted from Fig. 30 of Lipari et al.
2004d). We suggested that a probable explanation for the link between
the extreme Fe\,{\sc ii} emission and the extreme velocity OF is that
 both are associated to the interaction of the star formation processes
 and the AGN, that generate extreme explosive/HyN events.

\section{THE SPECTRAL EVOLUTION of COMPOSITE AGNs with OUTFLOW
(Theoretical Results)}
\label{spevol}

We propose to study the evolution on the interaction between a nuclear
SB, the SMBH and the OF.

\subsection{The evolution of NLR, associated with outflow/shells}
\label{nlrevol}

If most of the energy from the stellar winds and SN explosions 
from the central star forming region is thermalized, a hot
cavity in the  interstellar medium (ISM) will be created. 
This hot ``bubble'' will expand and shock the cluster ISM.
A region of constant high pressure
will form inside the radius that contains all the SN (r$_{SN}$
 Chevalier and Clegg 1985).

The bubble will expand faster along the steepest density gradient 
(i.e. along the poles in the case of a disk),
creating an elongated hot cavity. The shocked ISM will cool in a time scale
proportional to $V_{shock}$ and form a thin dense shell. 
The cooling time is,

$t_{cool}\sim 2000 {v_8 \over n_3} \ {\rm yr}$

and cooling distance,

$r_{cool}\sim 2 {v_8^2 \over n_3} \ {\rm yr}$

\noindent

where $v_8 = V_{shock}/ 10^8$ cm s$^{-1}$, and the pre-shock density is
$n_3=n_{interstellar}/10^3$ cm$^{-3}$.

The thin shell in pressure equilibrium with the hot cavity 
will reach a density

$n_{sh} = {P_0 \over T_{sh}} = 10^8 k_{-1}^{-2} T_{sh,4}^{-1}$

where $T_{sh,4} = T_{sh}/10^4 $K is the temperature of the thin shell.
This fast expanding thin shell photo-ionized by the radiation from
the core of the cluster will emit relatively
broad, FWHM $\sim V_{sh} \sim 2000$ km s$^{-1}$ forbidden and permitted 
lines. Among the forbidden lines stronger than H$\beta$ are
[O\,{\sc iii}]$\lambda$ 4363, 5007, 4959, [OI]$\lambda$ 6300. 

The continuous energy input of the SN will make the bubble to expand 
until it reaches
the edge of the cluster gas density distribution. 
At this point, the shock will accelerate and Rayleigh-Taylor
instabilities will set up and break the dense shell.
The hot thermal gas will escape between the fragments as a wind.
The ionizing radiation will also escape and start to ionize the
galactic ISM. Interaction of the out flowing wind with clouds
orbiting the bulge can give rise to radial motions and some X-ray
emission. The outfloing wind will accelerate the clouds.
The terminal velocity reached by dense clouds ($V_{cloud}$)
i.e. those where the wind can drive an isothermal shock  will be,

$V_{cloud} \sim n_{cloud}^{1/2} V_{wind}  $

where $n_{cloud}$ is the density of the fragments and $V_{wind}$ is the
wind velocity (Franco et al 1993). 
High density clouds are not much affected by the
wind while low density clouds and the ISM will be accelerated almost to 
half of the wind velocity.

After some time the wind will shock  the galaxy ISM at large distances 
from the nucleus. 
The size of the shock driven by the wind into the galactic ISM is
(Weaver et al. 1977, Chevalier \& Clegg 1985),

$R_{wind} \sim 4 \left ({\nu_{sn} \epsilon_{51} t_7^3} \over n \right )^{1/5}
Kpc$

More ionizing radiation will reach the galactic ISM as the fragmented
shell expands and finally disperse. This will giving rise to optical
filaments emitting narrow lines with line ratios typical of the NLR
and line width corresponding to the velocity field of the ISM of the
galaxy, i.e. $200 <$ FWHM $< 1000$ km s$^{-1}$
depending mainly on the mass of the galaxy.

The observed line ratios, FWHM and size of the NLR will therefore 
evolve on time scales
comparable to the time scale for the wind development. This time scale
will depend on the rate of energy input size of the SN region and
on the details of the gas distribution.
There will be differences induced by the shape of the star forming region.
Disk like regions of star formation will produce collimated winds while
by contrast spheroidal regions will produce spherical winds.

Finally,
it is important to remark that in the nearest BAL + IR + Fe\,{\sc ii} QSOs
--Mrk 231-- very deep 3D Gemini+GMOS spectroscopic data show
the absence of the NLR (L\'{i}pari et al. 2006). We suggested that
the multiple explosive events, detected previously in Mrk 231, could
explain --at least in part-- the absence of the NLR.
In addition, a similar result was found 
 in the other nearby BAL + IR + Fe\,{\sc ii} QSO:
IRAS 07598+6508 (Veron et al. 2006).

\subsection{The evolution of BLR, associated
with multiple compact SN remnants}\label{blrevol}

The composite QSO phase lasting from $\sim 8$ Myr to $\sim 60$ Myr 
is dominated by the type II SN activity and their
remnants. 
The SNR of metal rich intermediate mass stars ( M$ \sim 8$ to $25$
M$\odot$) are presumably very luminous because their kinetic energy
is rapidly thermalized by dense circumstellar material around the red
supergiant progenitor evolving in the high pressure ISM of the star
forming region.
These compact remnants are probably the BLR of some AGNs.

Supernova remnants evolving in a dense  and homogeneous CSM
($n > 10^5$~cm$^{-3}$ ) reach their maximum luminosity ($L >
10^7$ L$\odot$) at small radii ($R < 0.1$~pc ), soon after the SN
explosion ($t < 20$~yr) and while still expanding at velocities of more
than $1000$~km s$^{-1}$ (Shull 1980; Wheeler et al.  1980; Draine and Woods
1991; Terlevich et al. 1992). In these compact SNRs, radiative cooling
becomes important well before the thermalization of the ejecta is
complete, making the remnant miss the Sedov track. As a result, the
shocked matter undergoes a rapid condensation behind both the leading
and the reverse shocks. Two concentric, high-density, fast-moving thin
shells are then formed. The cool, dense shells, the freely expanding
ejecta, and a section of the still dynamically unperturbed interstellar
gas, are all irradiated and ionized by the photon field produced by the
radiative shocks.

Terlevich et al. (1992) showed that the predicted line intensity ratios
from the simple cSNR model
are in good agreement with those observed in the BLR of AGN. 
In addition the model has the following properties in common with the
BLR of AGN:
Peak bolometric luminosity over $10^{43}$ ergs,
emission line width of about $5000$ km s$^{-1}$,
BLR size of about $0.01 pc$,
two main broad line emitting regions, one with high density
($n_e \sim 10^{12}$ cm$^{-3}$)
and low ionization (LIL), and the other of lower density ($n_e \sim 10^{10}$
cm$^{-3}$)
and higher ionization (HIL),
column density of ionized gas of about $10^{23}$ cm$^{-2}$,
total mass of ionized gas in the BLR from about $1$M$\odot$ ~to
about $10$M$\odot$,
power-law ionizing spectrum of the form $f_\nu\propto\nu^{-0.5}$
up to $\sim 100 kev$  with a bump between $10$ eV and $400$ eV,
absence of broad forbidden lines,
stable BLR emitting gas,
small X-Ray absorption column density,
redshift and line-width differences between HIL and LIL systems.

Terlevich et al.  (1992, 1995) have analysed the time dependent
process that occurs prior
to thin shell formation in rapidly radiating cSNR during its maximum
luminosity. They have found  that an inherent delay between
the photon emission
and the time required to increase the density of the cooling gas, leads
to a lag between the observed continuum burst and the
emission lines response. This delay is intrinsic to the physical
processes investigated and not related to the geometry of the system;
there are in fact no light-crossing time arguments involved. The total
width of the cold or photoionized region is only about $10^{13}$ cm
(about 300 light seconds), yet, delays
of up to several weeks between continuum and lines emission are generated.

The initial conditions for
modelling cSNR in the work of Terlevich et al. (1992, 1995) 
were adequate for a low mass SN progenitor (M $\sim 7$ M$\odot$ )
that represents the most common type II SN event. More massive progenitors
will have different evolution involving
slower shocks due to more massive ejecta.
Also the ejecta composition may be very metal rich in massive type II SN.
Due to the low sock velocity and the large metal content reverse shock
cooling times will be very short and as a consequence
the evolution of the reverse shock may dominate the radiative 
phase of these remnants.

In young stellar populations Fe is produced exclusively by 
massive (M$>12$ M$\odot$) type II 
SN. SN type II with progenitors with initial masses below that limit produce
basically no Fe (see Renzini et al. 1993). Theoretical models of 
nuclesynthesis in massive type II SN
predicts the Fe/H ratio to be more than 10 and up to 30 times over-solar. 
Models also predict that only SN with progenitors more
massive than $12$M$\odot$ will produce Fe. 
This implies that during the evolution
of a stellar cluster there will be a well defined epoch during which the
ejecta from  SN will be Fe rich. This corresponds to cluster ages between
$\sim 8$ Myr, the life-time of a $25$M$\odot$ stars at the upper limit for
the mass of a type II SN progenitor, and $\sim 20$ Myr, the life time of a
$12$ M$\odot$ (interestingly this is the same age range than that of
red-Supergiants).

During the evolution of a cluster in the SNII stage 
the turn~off mass diminishes from $\sim 25$ M$\odot$ to $\sim 7$ M$\odot$.
At the same time the mass of the SN ejecta will diminish 
from $\sim 20$ M$\odot$
down to maybe $\sim 1$ M$\odot$. 
If the energy per SN does not change with the mass
of the progenitor, the ejecta of the most massive progenitors will be slower
than that of the lower mass ones by 
$\sim 20^{1/2} \sim 4.5$ assuming energy conservation.

The slower ejecta will drive slower and cooler shocks. The emitted
X-ray spectrum of massive cSNR is thus predicted to be deficient
in hard X~ray with respect to that of low mass normal cSNR.
Terlevich et al. (1992) found a turnover in the emitted spectrum
at around 20 to 40 kev. The simple scaling above would suggest
that for massive cSNR the turnover should be around 1 kev .
The estimate of the shape of the emitted x~ray spectrum by a population
of massive cSNR will have to wait  detailed computations of the
evolution of a massive cSNR.

Because of the slower reverse shocks and the much higher abundance
the evolution of the reverse shock is expected to dominate
the luminosity evolution of the remnant. In these conditions,
the main sources of ionizing radiation would be the reverse shock
while the reprocessing will be done by the ejecta and the 
reverse shock thin shell.

Therefore early in the evolution of a young cluster evolving in a high
pressure environment, at a time when the turn-off mass is between
25 and 12 M$\odot$, the cSNR will emit relatively narrow broad lines
with FWHM$\sim 2000$ km s$^{-1}$, strong Fe\,{\sc ii} lines and will be
very deficient in hard X~rays.
Later in the evolution of the cluster, when the turn-off mass falls below
$12$ M$\odot$ the cSNR emits broader lines with FWHM$\sim 5000 $km s$^{-1}$,
weak or no Fe\,{\sc ii} lines and strong hard X~rays.

\section{Composite BAL processes, and  the role of 
supergiant shells and hyper novae}\label{blrgchnr}

In section 3.4, we already noted that for Mrk 231
were found evidence that the BAL I and III systems
are probably associated with: the AGN sub relativistic jet and the
nuclear SB with giant explosions and expanding shells, respectively
(L\'{i}pari et al. 2005a; Punsly \& L\'{i}pari 2005). Thus, it
is important to study this type of composite BAL process/scenario.

In particular, 
the presence of multiple concentric expanding supergiant bubbles/shells
(in Mrk 231),
with centre in the nucleus and with highly symmetric circular shape could be
associated mainly with giant  symmetric explosive events (L\'{i}pari et
al. 2005a,c,d). These giant explosive events could be explained in a
composite scenario: where mainly the interaction between the starburst and
the AGN could generate giant explosive events.
In particular, Artymowicz, Lin, \& Wampler (1993) and Collin \& Zahn (1999)
already analysed the evolution of the star formation (SF) close to  super
massive black hole (SMBH) and inside of accretion disks. They suggested that
the condition of the SF close to the AGNs could be similar to those of the
early/first SF events, where giant explosive processes are expected,
generated by hypernovae (with very massive progenitors: M $\sim$100--200
M$_{\odot}$; see Heger \& Woosley 2002; Heger et al. 2003, 2002).
In accretion disk, the star--gas interactions can lead to a special mode of
massive star formation. Furthermore, the residuals of the first  SNe
(neutron stars) can undergo a new accretion/interaction phase, with the gas,
leading to very powerful SN or hypernova explosions.

Furthermore,
an explosive scenario for the origin of the BAL III system (in Mrk 231) could
explain: the shape of the light curve variability, and also the presence of
multiple concentric expanding superbubbles/shells (L\'{i}pari et al.
2005a,c,d).
Recently, one of the most important developments in the study of SNe is the
discovery of some very energetic SNe, whose kinetic energy exceeds
10$^{52}$ erg (called hypernovae, HyNe; Paczynsqui 1998; Wang 1999;
Galama et al. 1999; Hjorth et al. 2003).
In particular, these HyNe were detected mainly associated with starbursts
and long duration gamma-ray bursts (GRB).

For the evolution and explosion of  very massive population III (or
primordial) stars the results of theoretical models suggest that these
stars explode as giant--HyN with energies of 10$^{53}$ erg (Heger et al. 2002;
Heger \& Woosley 2002; Nomoto et al. 2004).
This giant--HyN with energies up to 100 times that of an ordinary
core collapse SN could be also an explanation for the origin of
superbubbles. 
Heiles (1979) already suggested that this single giant SN or HyN scenario
need to be considered (together with the multiple SN explosions model) 
in order to study the origin of supergiant bubbles. Thus,
this single/few giant--HyN scenario is a probable and interesting option
to consider for the origin of the expanding supershells, detected in
BAL + IR + Fe\,{\sc ii} QSOs.

For the origin of BAL systems in QSOs/AGNs, in the composite starburst+AGN
with OF/GW scenario, different theoretical models were proposed.
In particular:
(i) in SN ejecta (very close to AGN with galactic wind),
which are shock
heated when a fast forward shock moves out into the ISM (with a
velocity roughly equal to the ejecta) and a reverse shock accelerates
back and moves towards the explosion centre; the blue absorption lines arise
since SN debris moving toward the central source are slowed down
much more rapidly -by the AGN wind- than is material moving away
(Dyson et al. 1992; Perry \& Dyson 1992; Perry 1992); and
(ii) for IR dusty QSOs with OF/GW, in the outflowing gas + dust
material the presence of discrete trails of debris (shed by
individual mass-loss stars) could produce the BAL features (Scoville \&
Norman 1995; Scoville 1992).

These two models studied physical processes for small galactic-scale.
Now we consider the alternative of expanding supergiant shell, as the
origin of BAL systems (which is similar to the SN ejecta model but at
large scale).
This alternative  was already argued in order to explain the BAL system
in the IR + Fe II QSO IRAS\,07598+6508 (L\'{\i}pari 1994; see also Bond et
al. 2001; Guillemin \& Bergeron 1997).

Specifically, important  theoretical results were obtained by
Tenorio-Tagle et al. (1999), who studied a scenario based on the
hydrodynamics of superbubbles/galactic wind powered by massive starbursts
that account for different type of BAL systems detected in star-forming
galaxies (Kunth et al. 1998; Mas-Hesse et al. 2003).
These type of starburst models could explain mainly the BALs associated with
low velocity OF.
For IR + Fe\,{\sc ii} QSOs the OF processes show extreme velocity
and very explosive events (probably associated with the interaction
between the AGN and the evolution of massive star formation process
in the accretion nuclear regions; proposed by Collin \& Zahn 1999).
In this last scenario the presence of hypernova explosions are expected.

Therefore, in this supergiant shell scenario the physical processes could
be similar to those studied by Tenorio-Tagle et al. (1999), but the
deposition of kinetic energy in the ISM by extreme SBs (with HyN) is larger
than in standard SBs.

On the other hand, for IRAS 04505--2958,
the UV  spectra clearly show a BAL system  at
C {\sc iv}$\lambda$1549 emission line (see L\'{i}pari et al. 2005a).
For this BAL system was measured $\lambda =$ 1978.5 \AA, corresponding to
an ejection velocity of --1645 km s$^{-1}$.
This ejection velocity is --within the errors-- the same that the
previous value of OF obtained using the offset method:
--1700 km s$^{-1}$! (by L\'{i}pari et al. 2004d).
The offset method used for the study of  OF--candidates IR QSOs
(including IRAS 04505--2958) means that the H$\beta$ broad line
component  is blushifted in relation to the narrow one. Thus this result
obtained for IRAS 04505--2958 suggests that the optical low ionization BLR
and the BAL could be originated in the same OF process or supershell.

Different works even proposed that the broad line  {\it emission} region
could be associated with OF processes. In particular, these works suggest
that the BLRs are located in the OF of: accretion disks, the ejecta of SN
remnants, shocked clouds (by a nuclear galactic wind) around SN remnants,
extended stellar envelopes, etc (see for references/review Sulentic,
Marziani, \& Dultzin--Hacyan 2000; Terlevich et al. 1992;
 Dyson, Perry \& Williams 1992; Scoville \& Norman 1988).
Thus an interesting questions is: could an explosive event
generate a very unusual spectrum similar to that detected in
the nuclear region of AGNs like Mrk 231?.
Figure 14 in L\'{i}pari et al. (2005a)
shows the superposition of the spectrum of the unusual radio
SN type II--L 1979c (observed in 1979 June 26.18; Branch et al. 1981).
Only using colours we can distinguish each spectrum, since they are
almost identical.
A more constant OF (than  a single and standard SN) could
explain even the optical spectrum of some BLR.

In particular,
Lipari et al. (2005a) suggested that Fe II emission could originate
in warm regions obscured from direct ionizing UV photons, the obscuring
material being in the form of expanding shells. The giant explosive
events occurring from the evolution of very massive stars would produce
shock-heated material. This suggestion is in good agreement with our
recent finding, for the BAL + IR + Fe II QSOs IRAS 07598+6508,
 that the properties  of  the BLR are consistent mainly with collisional
 rather than radiative models (Veron et al. 2006).

Therefore, following the results discussed in the last two sections,
in the composite SB + AGN scenario even the emission of the BLRs could
be associated with one or a combination of the following physical processes:
(i) single or few HyN explosions  with OF shells,
(ii) multiple cSNRs,
(iii) standard BLR of clouds moving around the AGNs, and
(iv) in the OF (or in the surface) of AGN accretion disks.

\section{PREDICTIONS OF THE EVOLUTIVE AND COMPOSITE MODEL FOR AGNs.}
\label{resultspredic}

Based in the conclusions and results from the previous sections
we can now proceed to describe in a semi-quantitative way the
evolution of a nuclear cluster and BH.
The evolution of the properties of the BLR with time is 
one of decreasing Fe abundance and
increasing mass of the SMBH. In particular:

\begin{itemize}

\item
{\bf The NLR size, emission line width and profiles.}

The NLR would evolve from a very compact one with the size of few times
the size of the cluster to a fully developed NLR of kiloparsec size. 
The increase in size of the NLR should be reflected in some systematic
change in the line width and profiles of the forbidden lines with time.
Simple models of wind driven bubbles predict that early in the evolution
of the bubble the velocities of the shocked gas will be close to about
half of the average wind velocity while at
the end of the evolution the gas motions will be closer to those of the ISM.
The NLR line profiles should reflect this change in same degree.
Although no precise prediction can be made without detailed models,
from a qualitative point of view it is still possible to predict that
the broad [O\,{\sc iii}] profiles associated with the relatively strong
Fe\,{\sc ii} emitters should have shell-like profiles while narrower
[O\,{\sc iii}] profiles should be either flat top (disk-like) or Gaussian
(spherical-like)

\item
{\bf The BLR line width.}

The line width of the BLR should evolve with time as the SMBH increases
its mass. Thus a relation between line width of the BLR and strength of
Fe\,{\sc ii} is expected.
The relation is in the sense that young objects with very strong
Fe\,{\sc ii} emission will have narrower lines in the BLR
while older objects with weak or absent Fe\,{\sc ii} emission will
have fully developed broad permitted lines.

\item
{\bf The X~ray spectrum.}

As the ejecta velocity of the cSNR increases with the age of the cluster, 
shock velocities are also expected
to increase, consequently the shock temperature will
increase with cluster age.
Simultaneously the  Fe\,{\sc ii}/ H$\beta$ ratio will decrease.
Relatively old clusters ( with age $> 20$ Myr ) will have very weak
Fe\,{\sc ii} emission and hard X~ray spectrum while young clusters will
emit strong Fe\,{\sc ii} and have a softer X~ray spectrum.

\item
{\bf The Red Supergiant phase.}

Stellar evolution models predict that the contribution by RSG to the
total emitted spectrum of a metal rich stellar cluster will be maximum 
at ages $8 < t < 20$ Myr . During this time some stellar features like
the IR CaII triplet at $\sim 8500 \AA$ should be particularly strong
(Terlevich et al. 1991).

\end{itemize}

\section{BAL + IR + Fe\,{\sc ii} PHENOMENA AND  THE
EVOLUTIVE AND COMPOSITE MODEL.}
\label{balirfe2}

\begin{itemize}

\item
{\bf The Broad Absorption Line systems:}

Strong evidence of the composite nature of BAL systems
were found in Mrk 231. Thus,
the  large galactic-scale outflow with  superbubbles in BAL + IR
+ Fe\,{\sc ii} QSOs (L\'{i}pari et al. 2003, 2005a,c,d) is an
interesting alternative for the origin of "some" BAL systems in these
IR objects. We already noted that
Tenorio-Tagle et al. (1999) studied a scenario based on the
hydrodynamics of superbubbles/galactic wind powered by massive starbursts
that account for different type of BAL systems detected in star-forming
galaxies, with low velocity OF. We suggest that for IR + Fe\,{\sc ii} QSOs
the physical processes could be similar to those studied by Tenorio-Tagle
et al. (1999), but the deposition of kinetic energy (in the ISM) by extreme
SBs is larger than in standard SB.

The effects of the orientation of the line of sight
and dust obscuration could play a secondary role,
in the observation of BAL systems. Specifically,
the BALs could be more easy to detect (in young QSOs)
if the winds are observed  edge on.

It is important to note that the supergiant shells/arcs scenario (for the
origin of BALs) is in agreement with an interesting
observational result: the detection of a high fraction of supergiant shells,
arcs and rings, in low-ionization BAL + IR + Fe II QSOs (Lipari et al. 2003).

\item
{\bf The strong IR continuum emission:}

An interesting prediction of the starburst + AGN model is related with
luminous IR galaxies (Sanders \& Mirabel 1996) as the progenitors of QSOs.
As discussed by Terlevich \& Melnick (1988), the early HII phase (of the SB) 
are associated with large amount of dust. This is
due not only to the dust present in any star formation region but also 
to the large amount of dust synthesized by the most massive stars
during the $\eta$-Carinae phase before becoming WR stars.
During this phase up to 1 M$\odot$ ~of dust per evolved massive star may be 
injected into the core ISM, enshrouding most of the massive stars.
The emitted luminosity will be therefore, dominated
by the far infra-red (FIR) luminosity and these systems will presumably be
present in large numbers in the IRAS sample of galaxies with Starburst and
Seyfert 2 nuclei.

The HII and Seyfert 2 phase that precedes the QSO 
phase, lasts $8 \times 10^6$ years, or about $1/10$ of the duration of
the QSO phase.
During this early phase, the cluster luminosity is at its maximum, and
on average, about 4 times higher than during the QSO phase.
The QSO progenitor should therefore be a luminous and short lived 
(short compared with the QSO life-time) FIR source.
Comparative studies of the luminosity function of IRAS galaxies and QSOs
should show if IRAS galaxies are, at a given epoch, about $4$ times more 
luminous in bolometric units and about $10$ times less
frequent than QSO.
This simple prediction applies only to the case
of a coeval population; systems where the star formation time scale is 
longer will show a mixture of all four phases at all times and will
therefore probably look like a QSO for most of their bright phase.

\item
{\bf The strong Fe\,{\sc ii} intensity and the [O\,{\sc iii}]/H$\beta$
vs. Fe\,{\sc ii}/H$\beta$  anti correlation:}

Because the Fe abundance of the  ISM in the nuclear region decreases
systematically with time from several times over solar to solar value, 
the Fe\,{\sc ii} intensity is expected also to decrease with time.

The first few generations of massive type II SN will clear the dust from
the inner parts of the cluster. The combined action of many X~ray/UV
flashes of the cSNR, reaching each of them $10^{10}$ L$\odot$ and
lasting about one year would evaporate the dust in the central regions
of the cluster. As SN activity proceeds, therefore the flux of hard
X~ray and UV radiation reaching galactic ISM at large distances should
increase with time. This is a necessary
condition to produce luminous forbidden lines. The model predicts
that the NLR will develop as the same time as the Fe\,{\sc ii}
intensity decreases.

\end{itemize}

\section{COMPARISON BETWEEN THE OBSERVATIONS AND THE MODEL.}
\label{compari}

To assess the reality of the relations
and search for new ones, we have compiled from the literature 
measurements of Fe\,{\sc ii}$\lambda$4570 / H$\beta$ , FWHM H$\beta$ ,
[O\,{\sc iii}]/ H$\beta$  ,IR luminosity, X~ray spectral slope for nuclei
with measured Fe\,{\sc ii}$\lambda$4570 / H$\beta$. Our data base of
Fe\,{\sc ii} AGNs and QSOs emitters is based mainly in the data published by
Bososon \& Green (1992); Joly (1991); Wills et al. (1985, 1992);
Lipari et al. (2005a, 2004a,b,c,d; 1993, 1991); Zheng et al. (2002), and
others.
Although the sample is non-homogeneous with respect to the
quality of the data and measurement procedures it is nevertheless useful
for statistical purposes. For example several correlations reported in the
literature can be illustrated with this sample (see Figure 3).


A correlation not discussed in the literature is that of [O\,{\sc iii}] FWHM
and the relative intensity of Fe\,{\sc ii} (Fig. 2b).
We have explored this relation and found that the FWHM of [O\,{\sc iii}]
is smaller in weak Fe\,{\sc ii} emitters
(this behaviour is the opposite of that of the FWHM of H$\beta$).
This is consistent with the finding of Boroson \& Green (1992)
that the peak [O\,{\sc iii}] is better correlated with the relative
intensity of Fe\,{\sc ii} than the [O\,{\sc iii}]
relative intensity.

Because of differential extinction between the receding and 
approaching sides of the expanding NLR it is expected that the blue
wing of the [O\,{\sc iii}] line should be stronger than the red wing.
This should introduce an asymmetry in objects with broad and relatively
faint [O\,{\sc iii}].

The model predicts that the IR CaII triplet in emission should reach
maximum brightness at ages between 10 and 20 Myr. At the same time
the triplet in absorption should also reach maximum intensity. Because
of the large intensity of the IR CaII triplet in emission it is 
in principle difficult to detect an underlying absorption where each
line has at most 10\AA\ equivalent width. It was very surprising
for us when searching the literature a posteriory that the 
stellar absorptions have been indeed detected in many of the 
strong Fe\,{\sc ii} and CaII emitters (Persson 1988; van Groningen 1993).

Persson (1988) detected CaII triplet absorption in two of his
strong CaII emitters (Mrk 42, Mrk 6). 
Because the ratio of absorption line intensities is 1:9:5 while
the emission lines are on average about equal, the most affected 
emission will be the 8542 \AA\ line while the least will be 
the 8498\AA\ line. 
Mrk 42 shows inversion in the
profiles that can be attributed to narrow stellar absorption. The case
of Mrk 6 is simpler due to the fact that the BLR is shifted with to the 
blue with respect to both the NLR and the stellar absorptions. 
Evidence for absorption can be found in other of Persson strong Fe\,{\sc ii}
emitters like Mrk 1239, Mrk 766 and most
noticeably for its similarity with Mrk 42 in Mrk 684. 
This is better seen in the Persson high dispersion data (Persson fig 4)
where the left side shows the spectra of Mrk~42, Mrk~1239 and Mrk~6
all of them with clear indication that 8542\AA\ is the stronger
of the absorption features detected.

van Groningen (1993) in an analysis of the
spectra of three Seyfert 1 galaxies with strong CaII emission detected
strong IR CaII triplet absorption in Mk231. On the other two nuclei
he observed that the FWHM of the CaII lines
is larger than the rest of the broad lines while the width at zero intensity
is the same. In the analysis of the excess FWHM of the 
CaII lines van~Gronningen did not considered the influence of the underlying
stellar absorption that, as pointed by Persson (1988; see appendix)
any study of IR CaII line profiles should pay attention to the 
stellar absorption effects,
as it is clear that the deduced FWHM values will be overestimated unless
profiles are first corrected for absorption.

The simultaneous occurence of strong IR CaII broad emission and
strong IR CaII stellar absorption, is a very important prediction 
of the model that deserves to be further investigated.
It provides strong and independent evidence that the sequence of
Fe\,{\sc ii} strength is a sequence in stellar age and not in
aspect ratio. Further independent information may be obtained from
studies of the $2.3 \mu $ CO index that as the IR CaII triplet
is very strong in RSG.

\section{CONCLUSIONS.}\label{discon}

The following relations are not explained inside current Unified models
of AGN (where differences between types are due entirely to orientation
effects):
(i) Strong BAL + Fe\,{\sc ii} QSOs tend to be also strong IR sources.
(ii) FWHM of H$\beta$ vs. Fe\,{\sc ii}/H$\beta$ anti-correlated.
H$\beta$ is narrower in strong Fe\,{\sc ii} emitters.
(iii) X~ray spectral index vs. Fe\,{\sc ii}/H$\beta$ anti-correlated.
The X~ray spectrum slope is steeper  in strong Fe\,{\sc ii} emitters. 
(iv) [O\,{\sc iii}]/H$\beta$  vs. Fe\,{\sc ii}/H$\beta$ anti-correlated.
[O\,{\sc iii}] is weaker in AGN with strong Fe\,{\sc ii}.

In this paper we have explored an evolutionary and composite Unified scenario
involving super massive black hole and starburst  with outflow,
that seems capable of explaining most of the observational properties of AGNs.
Our suggestion is explored  inside the expectations of the Starburst model
close associated with the AGN.

In particular,
the prominence of BAL + Fe\,{\sc ii} emission in quasars with strong infrared to
optical luminosity ratio is consistent with the assumptions of a
composite heating source linked to violent SB and AGN processes
(L\'{i}pari et al. 1994, 2003, 2004a,b,c,d, 2005a).

The Fe\,{\sc ii} emission could be associated with rapidly cooling supernovae remnants
and/or expanding shells occurring in the central regions of galaxies
undergoing intense starburst activity (Terlevich et al. 1992; Lipari et al.
2005a; Veron et al. 2006).
We suggest that the observed properties of AGN with Fe\,{\sc ii} emission can be
understood as an evolutionary sequence where the observed differences between
strong and weak Fe\,{\sc ii} emitters and the observed correlations with
Fe\,{\sc ii} relative intensity, are related to evolutionary changes in the
cSNR and SN activity and the development of the NLR in a nuclear starburst.

Theoretical models of type II SN nucleosynthesis indicate that type II
SN with masses larger than 12 M$\odot$ 
synthesise large amounts of iron while stars of lower mass produce little
or no iron. Photoionization models of cSNR with ejecta abundances
corresponding to the theoretical predictions indicate that cSNR associated
with massive progenitors typical of young Starbursts (age about 10 Myr) 
will have very strong Fe\,{\sc ii} emission lines. 
The BLR of AGN with strong Fe\,{\sc ii} emission could be massive cSNR
corresponding to relatively young ($8$ Myr $<$ age $< 20$ Myr) Starbursts
with during massive ($\sim 25 > M > \sim 12$ M$\odot$) type II SN activity.
In these relatively young and dusty starburst the NLR is small and confined
to the proximity of the starburst therefore generating only weak forbidden
lines. The large ejected masses in these massive type II SN will also
produce cSNR with relatively lower ejecta velocity and therefore lower
velocity shocks, narrower permitted emission lines and softer X~ray
spectrum than those of low mass cSNR ($\sim 10 > M > \sim 7$ M$\odot$)
typical of older starburst ($40$ Myr $<$ age $< 60$ Myr) .
These older starbursts have low mass SN progenitors that synthesise
little or no Iron at all, and eject a smaller mass.
Thus, older starburst will have cSNR with weak Fe\,{\sc ii} lines,
high velocity shocks, broader emission lines and harder
X~ray spectrum.
Due to the thermal pressure of the SN activity the NLR expands reaching 
maximum development only in older starburst. Only these older
starburst will show bright forbidden lines and large, i.e. fully
developed NLR.

We find good agreement between the predictions of the model and the observed
correlations between Fe\,{\sc ii} intensity, BLR line width, [O\,{\sc iii}]
intensity and X~ray spectral slope of AGN. 
Additionally, predictions relating the Fe\,{\sc ii} intensity to the FWHM
and asymmetry parameter of the [O\,{\sc iii}] lines are corroborated.

Radio loud systems may be associated with the later stages of 
evolution. We also argue that the fact that radio loud AGN double
or extended morphology tend to show no Fe\,{\sc ii} and very broad permitted
lines while core dominated AGN tend to show Fe\,{\sc ii} in emission and
narrow permitted lines suggest also an evolutionary sequence.
Furthermore, since the timescale for the development of megapersec size
radio sources may be similar to the lifetime of the AGN (Fanti et al. 1995),
the  size of radio sources should not depend solely on projection effects.

We propose a check for this evolutionary model based in the
fact that the epoch of Fe rich ejecta ($8$ Myr $<$ age $< 20$ Myr)
coincides with the peak of Red Supergiant activity ($8$ Myr $<$ age
$< 15$ Myr) in metal rich stellar populations. Thus the underlying
stellar continuum of strong Fe\,{\sc ii} emitters should have stronger
IR CaII triplet absorption lines than that of weak Fe\,{\sc ii} emitters.
Because of the strong age dependence of the RSG activity, a statistical
study of the strength of the IR CaII triplet absorption as a function of
Fe\,{\sc ii} intensity can provide a powerful test of our model.

In Summary, in the evolutive and composite unification scenario 
the following main ideas were proposed (or confirmed):
(i) BAL QSOs are young systems with composite OF.
(ii) The Fe\,{\sc ii} intensity provides an age indicator for BLR AGNs.
(iii) BLR emission line FWHM changes with age.
(iv) X~ray slope changes with age.
(v) [O\,{\sc iii}] intensity and FWHM changes with age.
(vi) The strong IR continuum emission is associated with the composite
nuclear nature.
(vii) The Sy1/Sy2 and BLRG/NLRG segregation is mainly due to
orientation/obscuration by toroid.

\section*{Acknowledgments}

The authors thank  M. Bergmann, H. Dottori, M. Joly, E. Mediavilla,
D. Merlo, B. Punsly, S. Sanchez, L. Sodre Jr., M. Veron,
S. Vine, X. Xia, and W. Zheng for assistance and discussions.
This research was carried out using the NASA Extragalactic Data base (NED).
This work was supported in part by Grants  from Conicet and SeCyT
of UNC (Argentina).


%
%
%

\clearpage

\begin{figure*}
\vspace{12.0 cm}
\begin{tabular}{c}
\includegraphics{fig1.ps}\cr
\end{tabular}
\vspace{8.0 cm}
\caption {IR colour--colour diagram for IR  mergers/QSOs with galactic
winds (from Table 1 of L\'{i}pari et al. 2005a) and for standard QSOs
(from the PG sample; Boroson \& Green 1992). The black dashed line
depicts a probable evolutive path: from IR mergers with low velocity OF
to IR QSOs with extreme velocity OF. In the transition area between the
black body and power law regions we found a clear sequence of BAL
 + IR + Fe II QSOs (starting from Mrk 231 and IRAS 07598+6508,
 which are located very close to the black body region).
 }
\label{f1ircc}
\end{figure*}

\clearpage

\begin{figure*}
\vspace{12.0 cm}
\begin{tabular}{c}
\includegraphics{fig2.eps}\cr
\end{tabular}
\vspace{6.0 cm}
\caption {
Plot of the ratio Fe\,{\sc ii}$\lambda$4570 / H$\beta$ vs. Velocity (offset)
OF, for IR QSOs. The data were obtained from Lipari et al. (2004d): Table 8. 
}
\label{f2fe2of}
\end{figure*}

\clearpage

\begin{figure*}
\vspace{12.0 cm}
\begin{tabular}{c}
\includegraphics{fig3a.ps} \cr
\includegraphics{fig3b.ps} \cr
\end{tabular}
\vspace{6.0 cm}
\caption {
Fig. 3a. the relation between the ratio of the equivalent widths (EW) of
Fe\,{\sc ii}$\lambda$4570 / H$\beta$ vs. Peak of [O\,{\sc iii}]. The open
and fill squares are bright and faint QSOs, respectively.
Fig. 3b. the relation between the ratio of the equivalent widths  of
Fe\,{\sc ii}$\lambda$4570 / H$\beta$ vs. FWHM of [O\,{\sc iii}]. The
dashed lines show linear fits of the two set of data.
 }
\label{fig3ab}
\end{figure*}

\clearpage

\begin{figure*}
\vspace{12.0 cm}
\begin{tabular}{c}
\includegraphics{fig3c.ps} \cr
\includegraphics{fig3d.ps} \cr
\end{tabular}
\vspace{6.0 cm}
\addtocounter{figure}{-1}
\caption {
Fig. 3c. the relation between the magnitud M$_V$ vs. FWHM of [O\,{\sc iii}].
The open and fill squares are strong and weak Fe\,{\sc ii} emitters,
respectively.
Fig. 2d. the relation between the Fe\,{\sc ii}$\lambda$4570 emission vs.
FWHM of [O\,{\sc iii}]. The dashed lines show linear fits of the different
set of data.
}
\label{fig3cd}
\end{figure*}

\end{document}